\documentclass{emulateapj}
\usepackage{aas_macros}

\usepackage{apjfonts}
\usepackage{graphicx}
\usepackage{hyperref}
\usepackage{amsmath,amssymb,amsbsy}
\usepackage[usenames]{color}

\usepackage{bm}

\newcommand\lsim{\mathrel{\rlap{\lower4pt\hbox{\hskip1pt$\sim$}}
        \raise1pt\hbox{$<$}}}
\newcommand\gsim{\mathrel{\rlap{\lower4pt\hbox{\hskip1pt$\sim$}}
        \raise1pt\hbox{$>$}}}
\newcommand\propsim{\mathrel{\rlap{\lower4pt\hbox{\hskip1pt$\sim$}}
        \raise1pt\hbox{$\propto$}}}

\newcommand{\D}{\mathrm{d}}

\newcommand{\yr}{\,\mathrm{yr}}

\newcommand{\hr}{\,\mathrm{hr}}

\newcommand{\nHz}{\mathrm{nHz}}
\newcommand{\Hz}{\,\mathrm{Hz}}
\newcommand{\Gyr}{\,\mathrm{Gyr}}

\newcommand{\Msun}{\mathrm{M}_{\odot}}
\newcommand{\km}{\,\mathrm{km}}
\newcommand{\ns}{\mathrm{ns}}
\renewcommand{\sec}{\mathrm{s}}
\newcommand{\pc}{\mathrm{pc}}
\newcommand{\mpc}{\,\mathrm{mpc}}

\newcommand{\BH}{\star}

\newcommand{\G}{\mathrm{G}}
\newcommand{\C}{\mathrm{c}}
\newcommand{\SMBH}{\bullet}
\newcommand{\A}{\mathcal{A}}
\newcommand{\orb}{\mathrm{orb}}
\renewcommand\vec[1]{\ensuremath\bold{#1}}
\newcommand{\forb}{f_\mathrm{orb}}
\renewcommand{\fp}{f_\mathrm{p}}
\newcommand{\np}{n_\mathrm{p}}
\newcommand{\rp}{r_\mathrm{p}}

\begin{document}
\title{Mapping the Galactic Center with Gravitational Wave Measurements using Pulsar Timing}

\author{Bence Kocsis,\altaffilmark{1}, Alak Ray,\altaffilmark{2,3}, Simon Portegies~Zwart\altaffilmark{4}}

\affiliation{$^1$Harvard-Smithsonian Center for Astrophysics, 60 Garden Street, Cambridge, MA 02138,
{\small bkocsis@cfa.harvard.edu}}

\affiliation{$^2$ Harvard-Smithsonian Center for Astrophysics, 60 Garden Street, Cambridge, MA 02138 and $^3$ Tata Institute of Fundamental Research, Homi Bhabha Road, Mumbai 400005, India, {\small akr@tifr.res.in}}

\affiliation{$^4$ Leiden Observatory, Leiden University,
                P.O. Box 9513, 2300 RA Leiden, The Netherlands,
                {\small spz@strw.leidenuniv.nl}}

\begin{abstract}
We examine the nHz gravitational wave (GW) foreground of stars and black holes (BHs) orbiting SgrA$^*$ in the Galactic Center. A cusp of stars and BHs generates a continuous GW spectrum below 40\,nHz; individual BHs within $1\mpc$ to SgrA$^*$ stick out in the spectrum at higher GW frequencies. The GWs and gravitational near-field effects can be resolved by timing pulsars within a few pc of this region. Observations with the Square Kilometer Array (SKA) may be especially sensitive to intermediate mass black holes (IMBHs) in this region, if present. A $100\,\ns$--$10\,\mu$s timing accuracy is sufficient to detect BHs of mass $1000\,\Msun$ with pulsars at distance $0.1$--$1\,\pc$ in a 3 yr observation baseline. Unlike electromagnetic imaging techniques, the prospects for resolving individual objects through GW measurements improve closer to SgrA$^*$, even if the number density of objects increases inwards steeply. Scattering by the interstellar medium will pose the biggest challenge for such observations.
\end{abstract}

\vspace{\baselineskip}


\vspace{\baselineskip}

\section{Introduction}
There is a great ongoing effort to use pulsar timing arrays to detect gravitational waves (GWs) in the nHz
frequency bands. GWs, if present, modify the exceptionally regular
arrival times of pulses from radio pulsars.
Observations of a correlated modulation in the time of arrivals (TOAs) of pulses from a network of highly stable
millisecond pulsars (MSPs) across the sky can be used to detect GWs \citep{1979ApJ...234.1100D,2011MNRAS.414.3117V}.
Existing pulsar timing arrays (PTAs)
utilize the brightest and most stable nearby MSPs in the Galaxy.

At nHz frequencies, the GW background is expected to be dominated by cosmological
supermassive black hole (SMBH) binaries { \citep{1995ApJ...446..543R,2003ApJ...583..616J,2003ApJ...590..691W,2004ApJ...611..623S}}.
Only a few studies considered the GW signal
from nearby sources. \citet{2001ApJ...562..297L} showed that pulsar signals would be sensitive to
a putative SMBH binary in the Galactic Center with mass ratio $0.06$, however, such a binary would
have other dynamical consequences which are not observed \citep{2003ApJ...599.1129Y}.
Further, \citet{2004ApJ...606..799J} showed that the nearby extragalactic source
3C 66B does not contain a massive SMBH
binary, because PTAs do not observe the expected GW modulation.
{ \citet{1987MNRAS.225P..51B} and \citet{1996A&A...315..396D} examined if the
variable gravitational field of nearby stars could be detected using pulsar timing in the cores of globular clusters,
and similarly, \citet{2005ApJ...627L.125J} considered the possibility of detecting GWs from intermediate mass black hole
binary sources using pulsars in the cluster. }

In this paper, we examine the prospects for directly detecting the
GW foreground and gravitational near-field effects of the Galactic Center (GC) using pulsars in the vicinity of this region { \citep[see also][]{1994IAUS..165.....C}}.
We estimate the foreground (in contrast with the cosmological background of GWs) generated by the dense population of stars and compact objects (COs)
in the GC, including about 20,000 stellar mass black holes (BHs)
\citep{1993ApJ...408..496M,2006ApJ...649...91F}
and perhaps a few
intermediate mass black holes (IMBHs) of mass $10^3\Msun$ \citep{2006ApJ...641..319P}. As these objects are much more
massive than regular stars populating the GC, they segregate and settle to the core  of the central star cluster. The emitted GW signal falls in the nHz range, well in the PTA frequency band.
While this foreground signal may be faint at kpc distances in the Galaxy as
\citet{2004ApJ...606..799J} discussed, it may exceed the
GW background locally, near the GC. The present generation of PTAs, achieve an
upper limit of the characteristic stochastic GW background amplitude of the order $h_c \lesssim 6 \times 10^{-15}$ at $f\sim 1/\yr \sim 30\,\nHz$ \citep{2011MNRAS.414.3117V},
while the theoretical prediction is $h_c \sim 9 \times 10^{-16} (f \yr)^{-2/3}$
\citep[and see \S~\ref{s:stochasticbg} below]{2011MNRAS.411.1467K}.
We estimate the level of pulsar timing accuracy necessary
(i) to constrain the mass of IMBHs in the GC using pulsars in the GC neighborhood to a level better
than existing constraints, or (ii) to resolve the central cusp of stellar mass objects.

While a large population of pulsars is expected to reside in the Galactic Center \citep{2004ApJ...615..253P},
their detection is quite challenging. It requires high sensitivity ($\gtrsim 0.01\,\mu$Jy) at relatively high radio
frequencies ($\gtrsim 10\,$GHz) where the pulse smearing due to the scattering of the ISM is less severe
\citep{1998ApJ...505..715L}. Since pulsars
have a steep radio frequency spectrum,
$S_{\nu} \propto \nu^{-1.6}$ to $\nu^{-1.8}$ \citep{1998ApJ...501..270K}, the
high frequency of observation
makes them very faint and thus difficult to detect and time.
Note however, that \citet{2011MNRAS.tmp..948K} have successfully detected nine radio pulsars at a
frequency of 17 GHz, including the detection of a millisecond pulsar and a magnetar
with an indication that the spectral index may flatten above 10 GHz.
The future Square Kilometer Array is expected to find several thousand regular pulsars, and a few MSPs in the GC
\citep{2004NewAR..48.1413C,SKA,2009A&A...493.1161S,2010ApJ...715..939M,2011A&A...528A.108S}.
High time resolution surveys recently found 5 millisecond pulsars in
mid galactic latitudes
\citep{2011MNRAS.411.1575B}, 3 within $100\,$pc of the Galactic Center
\citep{2006MNRAS.373L...6J,2009ApJ...702L.177D}.
Based on the properties of nearby pulsars and nondetections in a targeted observation,
\citet{2010ApJ...715..939M} recently put an
upper limit of 90 regular pulsars within the central $1\,$pc. It is likely that the timing accuracy of these pulsars
in the GC will be much worse than those in the local Galactic neighborhood. However, the GW signal may be much stronger
near the GCs to make a GW detection possible with a lower timing accuracy.

We adopt geometrical units $\G=\C=1$, and suppress the $\G/\C^2$ and $\G/\C^3$ factors to
change mass to length or time units.

\section{Black Holes in the Galactic Center}
\subsection{Stellar mass BHs}
The number density of objects in a relaxed galactic cusp orbiting around an SMBH with semimajor axis $a$ is
\begin{equation}\label{e:nr}
 n_{\BH}(a) = \frac{3-\alpha}{4\pi} N_{\BH} \left(\frac{a}{\pc}\right)^{-\alpha} {\pc}^{-3}\,,
\end{equation}
where $N_{\BH}$ is the total number of objects within $1\,\pc$, and
$\alpha=7/4$ for a stationary Bahcall-Wolf cusp \citep{1976ApJ...209..214B,2008gady.book.....B}.
If the mass function in the cusp is dominated by heavy objects, the density  of light and heavy objects relaxes to a profile with
$\alpha\sim 3/2$ and
$\alpha\sim 7/4$, while in the opposite case as steep as $\sim 9/4 $ to $3$ for the heavy objects \citep{2009ApJ...697.1861A,2009ApJ...698L..64K}.
These stationary profiles have a constant inward flux of objects which are eventually swallowed by the SMBH
and replenished from the outside.
We  take the theoretically expected value $N_{\BH}=20,000$ for stellar mass black holes
of mass $m_{\BH}=10\,\Msun$ within $1\,\pc$
\citep{1993ApJ...408..496M,2000ApJ...545..847M,2006ApJ...649...91F} and assume $\alpha_{\BH, \rm BH}=2$.

The eccentricity distribution for a relaxed thermal distribution of an isotropic cusp is such
that the number of objects in a $\D e$ bin is proportional to $\D N \propto e\, \D e$, independent of semimajor axis
\citep{2008gady.book.....B}.

\subsection{Intermediate mass BHs}
Intermediate mass black holes (IMBHs) are expected to  be created by the collapse of Pop III stars in the early universe
{ \citep{2001ApJ...551L..27M}},
runaway collisions of stars in the cores of globular clusters \citep{2002ApJ...576..899P,2006MNRAS.368..141F},
or the mergers of stellar mass black holes \citep{2006ApJ...637..937O}.
If globular clusters sink to the galactic nucleus due to
dynamical friction, they are tidally stripped and deposit their IMBHs in the Galactic nucleus. Then the IMBHs settle to
the inner region of the nucleus by mass segregation with stars.
\citet{2006ApJ...641..319P} predict that the inner $10\,$pc
of the GC hosts 50 IMBHs of mass $M\sim 10^{3}\Msun$.

There are very few unambiguous observations of  intermediate mass black holes (IMBHs) in the
Universe \citep{2004IJMPD..13....1M}. Ultraluminous X-ray sources provide the best observational candidates.
In particular, HLX-1 in ESO 243-49 is found to have a mass between $3\times 10^3\Msun \lesssim M \lesssim 3\times 10^5\Msun$ \citep{2011ApJ...734..111D}. In the Galactic Center, $0.13\,\pc$ projected distance from SgrA$^*$, IRS 13E is a
dense concentration of massive stars, which has been argued to host an IMBH
of mass between $10^3$ and $10^4\,\Msun$ \citep{2004A&A...423..155M}, however the
observed acceleration constraints make an IMBH interpretation in IRS 13E
presently unconvincing \citep{2010ApJ...721..395F}.
Astrometric observations of the radio source SgrA* corresponding to the SMBH can be used
to place an upper limit of the mass of an IMBH to $M\lesssim 10^4\,\Msun$ in
$5$--500 mpc \citep{2004ApJ...616..872R}.
Further, an IMBH could have served to deliver the observed young stars in the GC \citep{2003ApJ...593L..77H,2009ApJ...695.1421F},
eject hypervelocity stars \citep{2003ApJ...599.1129Y,2005MNRAS.363..223G}, create a low-density core
in the GC \citep{2006MNRAS.372..174B},
efficiently randomize the eccentricity and orientations of the observed S-star orbits \citep{2009ApJ...693L..35M},
and may have contributed to the SMBH growth \citep{2006ApJ...641..319P}.
These dynamical arguments can be used to place independent limits on the existence and mass of IMBHs
in the Galactic Center \citep[see][for a review]{2010RvMP...82.3121G}.
Future observation of the pericenter passage of the shortest period known star S2
may improve this limit to a ${\rm few} \times 10^3\Msun$ in 2018 \citep{2010MNRAS.409.1146G},
and even better limits will be possible by imaging SgrA* with
Event Horizon Telescope (EHT), a millimeter/submillimeter very long baseline interferometer (VLBI)
\citep{2011ApJ...735...57B}. We examine whether pulsar timing could detect an IMBH with parameters
not excluded by existing observations, or be used to place independent limits.

\subsection{Loss cone}\label{s:losscone}

A depleted region is formed in phase space if objects are removed at a rate faster than they
are replenished from outside by inward diffusion. In this region, Eq.~(\ref{e:nr}) is no longer valid.
The dominant source of removing stars is tidal
disruption or physical collisions, while for BHs it is GW capture by the SMBH.

The objects with initial conditions $(a,e)$ fall in and merge with the SMBH due to GW emission in a time
\begin{equation}\label{e:tmg}
t_{\rm mg}=\frac{5\kappa }{256} \frac{a^4 }{m_{\BH} M_{\SMBH}^2} (1-e^2)^{7/2}
= 36\,\kappa m_3^{-1} a_{\rm mpc}^4 (1-e^2)^{7/2}\Gyr.
\end{equation}
where $a_{\rm mpc}=a/{\rm mpc}$, $m_3 = m_{\BH}/10^3 \Msun$,
and $1\leq \kappa < 1.8$ is a weakly dependent function of eccentricity
\citep{1964PhRv..136.1224P}. Assuming $e\gg 0$ and $t_{\rm d} \leq t_{\rm mg}$,
the minimum semimajor axis is
\begin{align}\label{e:alc}
 a_{\rm lc} &= 0.4 \mpc \times m_3^{1/4} (1-e)^{-7/8} t_{{\rm d},9}^{1/4}\,,
\end{align}
where the inward diffusion time is parameterized as $t_{{\rm d},9} = t_{\rm d} / 10^{9} \yr$,
and we assumed $f\sim \fp \sim (1-e)^{-3/2}\forb$ (see Eq. (\ref{e:fp})).
For stellar mass BHs, the inward diffusion time is related to two-body relaxation \citep{2008gady.book.....B}.
Depending on the number and masses of BHs, \citet{2009MNRAS.395.2127O} find that
$0.1 \lesssim t_{{\rm d},9} \lesssim 10$.
For IMBHs, the inward diffusion is due to dynamical friction and the scattering of
stars. This process is initially faster than the relaxation timescale, but then slows down ($t_{{\rm d}, 9}\sim 10$)
as stars on crossing orbits are ejected by the IMBH \citep{2009ApJ...705..361G}.
The number density inside $a_{\rm lc}$ is expected to be much less than that of Eq.~(\ref{e:nr}).
{ In such a state, the eccentricity of an IMBH is increased \citep{2007ApJ...656..879M,2010ApJ...719..851S,2011ApJ...731L...9I}.
 However, a possible triaxiality of the cluster might result in the refilling of the loss-cone and
shorter inward migration timescales \citep{2011ApJ...732...89K,2011ApJ...732L..26P,2012ApJ...744...74G}.
}

\section{Gravitational Waves from the Galactic Center}

We start by reviewing the essential formulas to derive the GWs generated by a population of binaries with circular orbits,
then turn to the general eccentric case. We discuss other details of the spectrum, regarding the
high frequency cutoff and splitting into discrete peaks,  at the end of the section.

\subsection{Unresolved circular sources}
The GW frequency for a circular orbit
is twice the orbital frequency,  $f=2 \forb$, and the corresponding orbital radius is
\begin{equation}\label{e:rf}
r(f) = M_{\SMBH} (\pi M_{\SMBH} f)^{-2/3} = 2.6 \, f_{8}^{-2/3}\mpc\,.
\end{equation}
where $f_{8}=f/(10^{-8}\Hz)$. In the last equality, the mass of the central SMBH in SgrA$^*$
is taken as $M_{\SMBH} = 4.3 \times 10^6 \; M_{\odot}$ \citep{2009ApJ...692.1075G}.
To put in context, $20\mpc$ (i.e. about 4,100 AU) is the distance to the observed S-stars, the innermost star, S2, has semimajor axis $4\mpc$ and pericenter $0.8\mpc$, and the stellar disk of massive young stars extends down to $30\mpc$ \citep{2010RvMP...82.3121G}.

The RMS strain generated by an object of mass
$m_{\BH}$ orbiting around a SMBH of mass $M_{\SMBH}$ on a circular orbit
at distance $D$ from the source in one GW cycle is
\begin{equation}\label{e:h0}
  h_0(f) = \sqrt{\frac{32}{5}} \,\frac{M_{\SMBH}m_{\BH}}{D\, r(f)}
  = 8.8\times 10^{-15} m_{3} D_{\rm pc}^{-1} f_{8}^{2/3}\,,
\end{equation}
where $D_{\pc}=D/{\pc}$, $m_{3}=m_{\BH}/(10^3 \Msun)$,
and the 0 index will stand for zero eccentricity.
The $\sqrt{32/5}$ prefactor accounts for RMS averaging the GW strain over orientation.

The GW strain of many independent sources with the same frequency but random phase adds quadratically.
For a signal observed for time $T$, the spectral resolution is $\Delta f = 1/T$.
Therefore the number of sources with overlapping frequencies is
\begin{equation}\label{e:dN}
\Delta N =   \frac{dN}{d r}\left|\frac{d r}{d f}\right| \frac{1}{T}.
\end{equation}
The number of GW cycles observed in time $T$ is $f T$.
The integrated GW signal with frequency $f$ is
\begin{equation}\label{e:hccirc0}
 h_c^2 = \Delta N\, (f T) h_0^2 = \frac{dN}{d r} \left|\frac{d r}{d f}\right|  f h_0^2
= \frac{dN}{d \ln f}  h_0^2,
\end{equation}
also called \textit{characteristic spectral amplitude} \citep{2001astro.ph..8028P}.
Therefore the characteristic GW amplitude from a population of sources
can be interpreted as the RMS GW strain in a logarithmic frequency bin.
Here $dN/dr$ is the number of sources in a spherical shell and
$dr/df$ is to be evaluated using Eq.~(\ref{e:rf}).
Note that the final formula (\ref{e:hccirc0}) is independent of $T$.\footnote{This is true
as long as the source distribution is effectively continuous in frequency space, so that individual sources
are unresolved, but not for resolved discrete sources (see Eq.~(\ref{e:hci}) and \S.~\ref{s:smallnumber} below).}

We can express $h_c$ with the number density
of objects using $dN/dr = 4\pi r^2 n_{\BH}(r)$, as
\begin{equation}\label{e:hccirc}
 h_c^2(f) =  \frac{8\pi}{3} r^3 n_{\BH}\hspace{-1pt}(r)\, h_0^2 =
\frac{256\pi}{15} \frac{M_{\SMBH} m_{\BH}^2}{D^2} \, (\pi M_{\SMBH} f)^{-2/3} \, n_{\BH}\hspace{-1pt}[r(f)]\,.
\end{equation}
Eq.~(\ref{e:hccirc}) gives the root-mean-square (RMS) GW signal level drawing the stars or BHs
from a density profile $n_{\BH}(r)$. { The equation shows that the spectral shape is different from $f^{-2/3}$,
describing the RMS cosmological background \citep{2001astro.ph..8028P}. }Note that $h_c$ is proportional to the RMS mass
of objects which may exceed the mean for a multimass population.
The GW signal for any one realization is well approximated by the RMS if $\Delta N\gg 1$.
Close to the center (which corresponds to high $f$), $\Delta N$ is small,
and the GW spectrum becomes spiky, which we discuss in \S~\ref{s:smallnumber} below.

The GW foreground may be different from the above estimate for sources with
significant eccentricity, for sources on unbound orbits, or if they are non-periodic or
evolve secularly during the observation, or if the population is anisotropic.
We elaborate on the corresponding effects on the GWs in the subsections below,
and discuss the characteristic GW frequency where the spectrum would be
affected by the loss cone and small number statistics.

\subsection{Eccentric periodic and burst sources}
Here we summarize the modification of the above estimates due to eccentricity, and
refer the reader to Appendix~\ref{app:eccentric} for details.

For eccentric sources, the GW spectrum of individual sources is no longer peaked around $f=2f_{\orb}$.
For mildly eccentric sources ($e\lesssim 0.3$), the GW spectrum is spiky, with
discrete upper harmonics, $f = n f_{\rm orb}$, with decreasing amplitude for $n>2$.
The observed width of each spectral peak is $\Delta f = 1/T$.
For larger eccentricities, the upper harmonics are stronger than the $n=2$ mode, and the
peak frequency corresponds to the inverse pericenter passage
timescale $f_{p}$, where $90\%$ of the GW power is between $0.2 f_{p} < f < 3 f_{p}$.

The GWs are in the PTA frequency band if the pericenter passage timescale is less than
the observation time, $T$.
We distinguish two types of sources: {\it periodic} and {\it GW burst sources}.\footnote{There
are also ``repeated burst'' sources which are highly eccentric
with orbital periods less than $T$, which satisfy Eq.~(\ref{e:aper}) \citep{2011arXiv1109.4170K}.
We do not distinguish repeated bursts from periodic sources here.}
Periodic sources have orbital periods shorter than $T$, or semimajor axis
\begin{equation}\label{e:aper}
 a_{\rm per} \leq 5.7\, T_{10}^{2/3} \mpc\,.
\end{equation}
At relatively low frequencies where small number and loss cone effects do not kick in,
these sources generate a continuous GW spectrum which is similar to that given by the circular formula
Eq.~(\ref{e:hci}) within a factor $2$, assuming an isotropic thermal eccentricity distribution
(see Eq.~(\ref{e:hc2}) and (\ref{e:Kper}) in Appendix~\ref{app:eccentric}).

Sources with semimajor axes larger than Eq.~(\ref{e:aper})
generate a GW burst during pericenter passage in the PTA band if their pericenter distance at close
approach is less than Eq.~(\ref{e:aper}). Only a small fraction of these burst sources contribute
to the PTA measurements, the ones which are within time $T$ from pericenter passage along their
orbits during the observation.
In Appendix~\ref{app:eccentric}, we show that the stochastic background of burst sources
is expected to be typically much weaker than the periodic sources, unless there are IMBHs on
wide eccentric orbits.

\subsection{Individually resolvable sources}\label{s:smallnumber}
The GW foreground generated by a population of objects is smooth if the average number per $\Delta f$
frequency bin satisfies $\langle \Delta N\rangle \gg 1$. If the number density follows Eq.~(\ref{e:nr}),
and the total number within $1\pc$ is normalized as $\bar{N}_{\BH}=N_{\BH}/(2\times 10^4)$, then
the spectrum becomes spiky
($\langle \Delta N\rangle \leq 1$) above
\begin{equation}\label{e:fsn}
f_{\rm res} = 4.2 \times 10^{-8} \Hz \times
10^{9(\alpha -2)/(9-2\alpha)}
\left[\frac{(3-\alpha)\bar{N}_{\BH}}{T_{10}}\right]^{3/(9-2\alpha)}
\end{equation}
where $T_{10}=T/10\yr$, and we used Eqs.~(\ref{e:rf}) and (\ref{e:dN}) for circular orbits.
The corresponding orbital radius is
\begin{equation}\label{e:rsn}
r_{\rm res} = 1.0 \mpc \times
10^{6(\alpha -2)/(2\alpha-9)}\left[\frac{T_{10}}{(3-\alpha)\bar{N}_{\BH}}\right]^{2/(9-2\alpha)}
\end{equation}
Sources within $r_{\rm res}$ generate distinct spectral peaks above frequency $f_{\rm res}$.
We refer to these sources as {\it resolvable}.

The total number of resolvable sources is
\begin{align}
 N_{\rm res} &= \int_0^{r_{\rm res}} n_{\BH}(r)\, 4\pi r^2\D r = N_{\BH} \left(\frac{r_{\rm res}}{\rm pc}\right)^{3-\alpha}\\
 &=
20 \times 10^{9(\alpha -2)/(9-2\alpha)}
\bar{N}_{\BH}
\left(\frac{T_{10}}{(3-\alpha)\bar{N}_{\BH}}\right)^{2(3-\alpha)/(9-2\alpha)}\,.
\nonumber
\end{align}
Note that $f_{\rm res}$, $r_{\rm res}$, and $N_{\rm res}$ are exponentially sensitive to the density exponent:
in a 10 year observation of $N_{\BH}=20,000$ BHs,
$f_{\rm res}=(19, 42, 110)\,\nHz$,
$r_{\rm res}=(1.7, 1.0, 0.53)\mpc$, and
$N_{\rm res}=(7, 20, 70)$
for $\alpha=(1.75, 2, 2.25)$, respectively.
Similarly, for $N_{\BH}=50$ IMBHs,
$f_{\rm res}=(2.0, 3.8, 6.4)\,\nHz$,
$r_{\rm res}=(7.6, 5.1, 3.5)\mpc$, and
$N_{\rm res}=(1, 3, 12)$ for $\alpha=(2.25, 2.5, 2.75)$, respectively.
If the supply of objects by two body relaxation is very slow, $t_d \gg 1 \Gyr$,
and the loss cone is empty (see \S~\ref{s:losscone}),
then the number of resolvable sources can be much less.
{\it The GW spectrum transitions
from continuous to discrete inside the PTA frequency band,
and the expected number of resolvable sources is typically non-negligible.}

For $f>f_{\rm res}$, the GW spectrum is $h_c(f)= 0$, except for distinct frequency bins
which include one resolvable source each. For the latter, the net GW signal amplitude in time $T$ is
\begin{equation}\label{e:hci}
h_{c,1}(f) = (f T)^{1/2}\, h_0(f) = 1.6\times 10^{-14} m_3 D_{\rm pc}^{-1} T_{10}^{1/2} f_8^{7/6}\,.
\end{equation}
Therefore $h_{c,1}(f)$ increases quickly for individual sources with decreasing orbital radius or
increasing frequency. The prospects for detecting
individual sources closer than $r_{\rm res}$ to SgrA* improves because the
GW amplitude increases, while the confusion noise per frequency bin
decreases { (see Eq.~(\ref{e:dt}) below for the corresponding timing residual)}.

Note that Eq.~(\ref{e:hci}) corresponds to the GW polarization averaged
integrated signal\footnote{More precisely, for circular sources,
$h_c^2 = \langle\int_0^T (h_{+}^2 + h_{\times}^2) f \D t \rangle$, where $\langle.\rangle$ corresponds
to averaging over source inclination. Note that throughout this manuscript
$h_c(f)$ is in dimensionless units, counts per frequency bin.} for stationary circular orbits.
We derive $h_c(f)$ for eccentric periodic and burst sources in Appendix~\ref{app:eccentric}.
In that case, the GW power of each source is distributed over many harmonics,
making the detection of individual sources potentially more challenging. However,
GR pericenter precession leads to an overall amplitude modulation of the
relative weights of the two GW polarizations \citep{2006PhRvD..73l4012K}. This effect increases inward,
which may improve the prospects for separating individual sources.

\subsection{Maximum GW frequency}
What is the maximum GW frequency for objects outside of the loss-cone, for which the
GW merger is still longer than the relaxation timescale?
The maximal emitted GW frequency corresponding to Eq.~(\ref{e:alc}) is
\begin{align}\label{e:fmax}
 f_{\rm lc} &= 1.6 \times 10^{-7} \Hz\times m_3^{-3/8} (1-e)^{-3/16}t_{{\rm d},9}^{-3/8}\,,
\end{align}
where note that the dimensionless diffusion time is $0.1 \leq t_{{\rm d},9} \leq 10 $ for multimass two-body relaxation models
(see \S~\ref{s:losscone}), and we assumed $f\sim \fp \sim (1-e)^{-3/2}\forb$ (see Eq. (\ref{e:fp})).
The GW spectrum is expected to terminate above $f_{\rm lc}$. Eq.~(\ref{e:fmax}) implies that
$10\,\Msun$ BHs generate a GW background with a maximum frequency of around $10^{-6} \Hz=(12\,{\rm day})^{-1}$
for mildly eccentric sources, and somewhat larger for more eccentric orbits.
For IMBHs with  $m=10^3\,\Msun$, $t_d\sim 10\Gyr$, and $e=0.9$, the corresponding cutoff frequency  is around $2 \times 10^{-7} \Hz=(60\,{\rm day})^{-1}$. {\it Sources outside the loss cone are in the PTA GW frequency band.}

\subsection{Frequency evolution}
The above treatment is valid only if the frequency shift is negligible
during the observation, which we discuss next.

The timescale on which the GW frequency evolves, is $f/\dot{f} \sim  t_{\rm d}$, where
$t_{\rm d}$ is the timescale on which sources drift inwards.
The frequency drift is then $\dot{f} T \sim f T/t_{\rm d}$. Compared to
the frequency resolution $\Delta f = 1/T$ during the observation,
\begin{equation}
 \frac{\dot{f} T}{\Delta f} \sim \frac{f T^2}{t_{\rm d}} \sim 3\times 10^{-8}\,  f_8 T_{10}^2 t_{{\rm d},9}\,.
\end{equation}
The frequency drift is negligible due to GW emission or two-body relaxation
for sources outside the loss cone.

\section{Gravitational Waves and Pulsar Timing}
The cosmological GW background and its detectability with PTAs can be summarized as follows.

\subsection{Stochastic cosmological background}\label{s:stochasticbg}

The cosmological GW background constitutes an astrophysical noise for measuring the GWs
of objects orbiting SgrA* in the GC.
At these frequencies the stochastic GW background is expected to be dominated by SMBH binary inspirals,
with a characteristic amplitude roughly \citep{2011MNRAS.411.1467K}
\begin{equation}\label{e:hcs}
 h_{c,\rm{s}}= 1.8 \times 10^{-15} f_{8}^{-2/3}\,.
\end{equation}

The GW background may be suppressed below the stochastic level of Eq.~(\ref{e:hcs})
by a factor 2--4 due to gas effects at $f_{8}<1$
\citep{2011MNRAS.411.1467K}, and because of small number statistics at $f_{8}>1$ \citep{2008MNRAS.390..192S}.
On the other hand, individual cosmological SMBH binary sources may stick out above this
background by chance if they happen to be very massive and relatively close-by \citep{2009MNRAS.394.2255S}.

\subsection{Sensitivity level of PTAs}\label{s:ptas}
If pulsars are observed repeatedly in $\Delta t$ time intervals for a total time $T$,
the spectral resolution is $\Delta f = 1/T$, and the observable frequency range is $1/T \leq f\leq 2/\Delta t$.
For $T=10 \yr$, $\Delta t= 1\,$week, therefore $3\times 10^{-9}\Hz \leq f\leq 3\times 10^{-6}\Hz$.

To derive the angular sensitivity of PTAs, let the unit binary orbit normal, the vectors pointing from
the Earth toward the pulsar,
and from the binary to the pulsar be respectively,
$\hat{L}^a=(\sin\iota \cos\lambda, \sin\iota \sin\lambda, \cos\iota)$,
$\hat{p}^a=(\sin\mu \cos\psi,\sin\mu \sin\psi,\cos\mu)$
and $\hat{k}^a=(0,0,1)$.
In the far-field radiation zone of a circular binary,
the metric perturbation is $h_{ab} = h^{+}_{ab}(t) + h^{\times}_{ab}(t)$, where
\begin{align}
h^{+}_{ab}(t) &= 2[1 +  (L^{c} k_c)^2]\, \A \sin\Phi(t)\, e^{\times}_{ab}\,,\\
h^{\times}_{ab}(t)&= 4\, L^{c} k_c\, \A \cos\Phi(t) \, e^{+}_{ab},\quad
\end{align}
where $\hat{k}_b$ gives the GW propagation direction, $\Phi(t)=2\pi f t$ is the GW phase,
and $\A = M_{\SMBH} m_{\BH} / (D\, r)$ is the GW strain amplitude without constant prefactors,
which we assume is much larger at the pulsar than at the Earth.
Note that $D$ is the distance
between $M_{\SMBH}$ and the pulsar, while $r$ is the distance between $M_{\SMBH}$ and $m_*$.
We orient the coordinate system so that
$e^{\times}_{ab}=\hat{e}^x_{a}\hat{e}^y_{b}+\hat{e}^x_{b}\hat{e}^y_{a}$ and
$e^{+}_{ab}=\hat{e}^x_{a}\hat{e}^x_{b}-\hat{e}^y_{a}\hat{e}^y_{b}$ with
$\hat{e}^x_{a}=(1,0,0)$,
$\hat{e}^y_{a}=(0,1,0)$.
Assuming that the pulsar is much closer to the source than the Earth, the TOAs of pulses
change as
\begin{equation}\label{e:timres}
\delta t = \int z(t) dt,
\end{equation}
where $z(t) \equiv [\nu(t)-\nu_0]/\nu_0=z_+(t) + z_\times(t)$,
where $\nu(t)$ is the observed pulsar frequency, $\nu_0$ is its unperturbed value (i.e. neglecting GWs), and
\begin{align}\label{e:z+}
 z_+ &= \frac{1}{2} \frac{\hat{p}^a \hat{p}^b }{1+ \hat{p}^a k_a}  h^{+}_{ab}
=(1- \cos\mu) \cos 2\psi\, (1+\cos^2 \iota) \A \cos\Phi\\\label{e:zx}
 z_\times &= \frac{1}{2} \frac{\hat{p}^a \hat{p}^b }{1+ \hat{p}^a k_a}  h^{\times}_{ab}
=(1- \cos\mu) \sin2\psi\, (2\cos \iota) \A \sin\Phi
\end{align}
\citep{1979ApJ...234.1100D,2009PhRvD..79b2002F,2009PhRvD..80h7101C}.
{ Note that unlike for cosmological sources the Earth-term is not included because it is negligible 
in the timing residual for nearby pulsars.}
Averaging the redshift factor over an isotropic distribution of $\hat{p}_a$ and $\hat{L}_a$,
and over one GW cycle,
\begin{equation}\label{e:zGW}
z_{\rm GW} \equiv \sqrt{\langle z_+ ^2\rangle + \langle z_\times ^2\rangle}
=\frac{4}{\sqrt{15}}\,\frac{ M_{\SMBH} m_{\BH}}{D\, r(f)} = \frac{1}{\sqrt{6}} h_0(f)\,,
\end{equation}
where $h_0(f)$ is the RMS GW strain defined in Eq.~(\ref{e:h0}).
The variations in the TOAs are given by the integrated change, $z_{\rm GW}/{2\pi f}$.
For an observation time $T$, the sensitivity increases with the number of observed cycles as $(f T)^{1/2}$,
\begin{equation}\label{e:dt}
\delta t_{\rm GW} =
\frac{z_{\rm GW}}{2\pi f} (f T)^{1/2} = \frac{1}{\sqrt{6}}\frac{h_c(f)}{2\pi f} = 100\,m_3 D_{\pc}^{-1}  T_{10}^{1/2} f_8^{1/6} \ns\,.
\end{equation}
Thus, instead of the integrated strain $h/(2\pi f)$, the timing SNR is proportional to $h_c/(2\pi f)$, which incorporates
the $(fT)^{1/2}$ factor.
In the last equality, we assumed the contribution of a single source only, Eq.~(\ref{e:hci}), but in general $h_c$
can represent many sources per bin, given by Eq.~(\ref{e:hccirc}).
Conversely, for fixed $\delta t$ timing accuracy, the detectable characteristic GW amplitude for an
average binary orientation and pulsar direction, is
\begin{equation}\label{e:hcPTA}
h_{c,\rm PTA} (f) = \sqrt{6}\, (2\pi f )\,\delta t
=1.1\times 10^{-17} \delta t_{10} f_{8}\,,
\end{equation}
independent of $T$, where $\delta t_{10}=\delta t/10\,\ns$. The sensitivity is better in certain orientations: by an average factor
$\sqrt{3}$ if the binary-pulsar-Earth angle is $90^\circ$ (with random binary inclination),
by $\sqrt{5/2}$ if the pulsar is on the binary orbital axis (with random pulsar-Earth direction),
and by $\sqrt{3}\times\sqrt{5/2}=2.7$ if both conditions hold.

{ As of 2009, 13 pulsars were known with timing accuracy better than $\delta t \sim 250\,{\rm ns}$, the best ones
reaching $20$--$30\,{\rm ns}$ \citep{2010CQGra..27h4013H}. 
The huge collecting area of the future Square Kilometer Array (SKA) implies that it will be
able to time many hundreds of pulsars at the level currently achieved for a few pulsars, and will perhaps find ones with much
better characteristics \citep[see also][]{2009arXiv0909.1058J}. 
However, the timing accuracy may be much worse for fainter sources further away (see \S~\ref{s:discussion} below).}

%%%%%%%%%%%%%%%%%%%%%%%%%%%%%%%
\begin{figure*}
\centering
\mbox{\includegraphics[width=8.7cm]{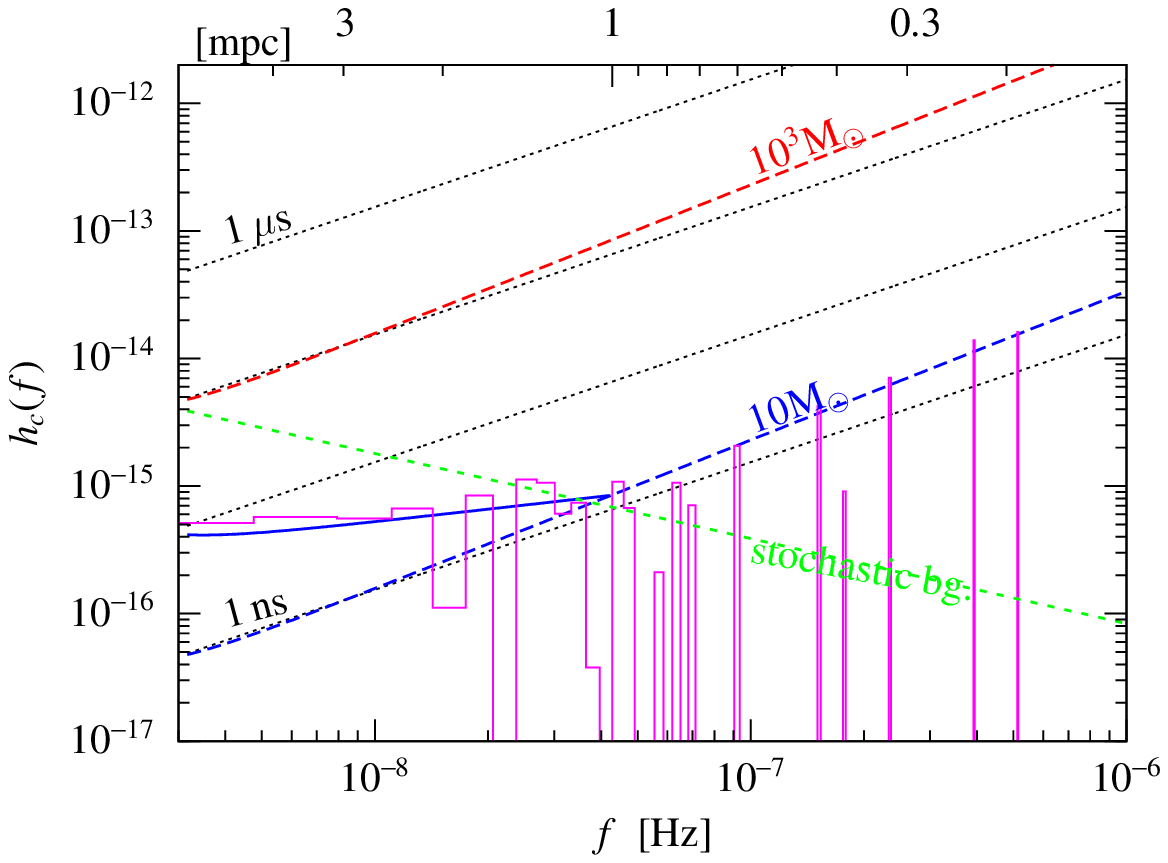}}
\mbox{\includegraphics[width=8.7cm]{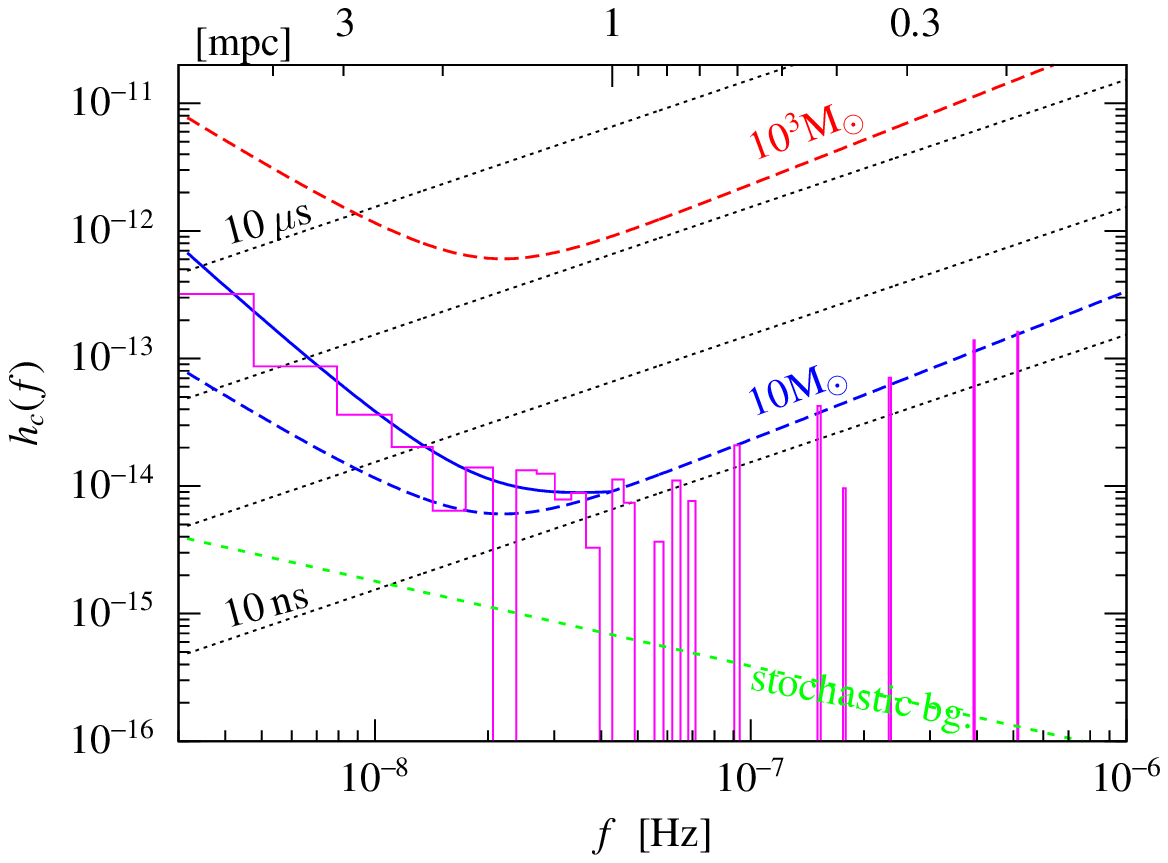}}
\caption{\label{f:hchar}
The characteristic strain amplitude $h_c$ vs. GW frequency $f$
for a pulsar at 1\,pc (left) and 0.1\,pc (right) from the Galactic Center.
Dotted black lines show the orientation-averaged $h_c$ for fixed timing residuals.
Blue and red dashed lines show respectively the orientation-averaged timing residuals
caused by individual stellar BHs and IMBHs on circular orbits
in a 10 yr observation, assuming 1 object per frequency bin.
Magenta lines show the timing residuals for a random realization of $10\Msun$
stellar BHs in the cluster { (20,000 BHs within 1 pc with number density $\propto r^{-2}$)}.
Typically only few bins are occupied at high $f$, generating a spiky signal with distinct sources.
At lower $f$, many sources overlap to create a continuous spectrum. The RMS average over
many realization of the cluster is shown by solid blue. The cosmological RMS stochastic GW
background (green dashed) can be much smaller.
%\vspace{0.1cm}
}
\end{figure*}

\subsection{Near-field effects}

Up to this point, we assumed that the pulsar signal is modulated by the leading order radiative GW terms,
and neglected other post-Newtonian near-field effects.
This is justified if the binary to pulsar distance is $D_p\gg \lambdabar$, where
\begin{equation}
\lambdabar = \frac{1}{2\pi f} = 0.16\, f_8^{-1}\, \pc\,.
\end{equation}
is the reduced GW wavelength. Thus, near-field effects will be negligible for $f_8>1$ for $D_p\gtrsim 1\,\pc$,
but they are significant at lower frequencies for pulsars closer to the center.

Let us estimate the order-of-magnitude of the leading order near-field corrections for the
particular geometry of this problem. The
variations in the TOAs are approximately determined by the metric $g_{\mu \nu}$ at the pulsar.
The near-zone metric induces a variation in the pulsar position and velocity, and modulates
the propagation of pulses \citep{2005ApJ...627L.125J}. We resort to simple
order-of-magnitude estimates neglecting possibly important lensing effects
\citep{1996A&A...311..746W,2009PhRvD..79b2002F}.
In the post-Newtonian (PN) approximation, the metric is practically a power series in
$m_i/r_j$, $r_i/r_j$, and $v_i$,  where $m_i$ are the masses
(i.e. $M_{\SMBH}$ or $m_{\BH}$),
and $r_j$ are distance parameters (i.e. $r$ or $D$), and
$v_i$ are the velocities \citep{1998PhRvD..58l4002B,2000PhRvD..61l4013A,2009PhRvD..80l4039J}.
The constant coefficients in this power series are typically order 1.
We look for periodically varying terms in the 2PN metric of the binary which might be comparable to $h_0(f)$ at the pulsar, and
assume the corresponding correction to the gravitational redshift and Doppler shift,
collectively called Einstein delay, is proportional to these terms.

For this estimate we restrict to circular binary orbits, where $m_{\BH}\ll M_{\SMBH} \ll r \ll D$,
with $m_{\BH}\sim 10^3 \Msun$, $r\sim \rm mpc$, $D\sim \pc$.
In this regime,
\begin{align}
 \frac{r}{D} &\sim 2.6\times 10^{-3} f_8^{2/3} D_{\pc}^{-1}\,,\quad
v \sim \sqrt{ \frac{M_{\SMBH}}{r}} = 8.5\times 10^{-3} f_8^{1/3}
\,,\nonumber\\
\frac{m_{\BH}}{r} &\ll \frac{M_{\SMBH}}{D}\ll \frac{M_{\SMBH}}{r} \sim  7.3 \times 10^{-5} f_8^{2/3}\,.
\label{e:approx}
\end{align}
We are not interested in terms that are not time-varying
such as $M_{\SMBH}/D$ or $M_{\SMBH}^2/D^2$, since these generate a constant
gravitational redshift. Remarkably, the mass dipole terms
${m_{\BH} r_{\BH}}/{D^2}$ and ${M_{\SMBH} r_{\SMBH}}/{D^2}$ are much larger
than the standard radiative term ${m_{\BH} M_{\SMBH}}/{r D}$ at these distances,
but these terms cancel out in the center of mass frame and are not present
in the metric.  (Here $\vec{r}_{\SMBH}=M_{\SMBH} \vec{r}/(M_{\SMBH} + m_{\BH})$
and $\vec{r}_{\BH}=m_{\BH}\vec{r}/(M_{\SMBH} + m_{\BH})$ give the vectors to the binary components
relative to the barycenter,  $\vec{r}$ is the binary separation, $\vec{D}$ is a field point.)
However, the pulsar perceives the binary
components at their retarded positions and so the dipole terms do not cancel out exactly unless the binary
line-of-sight velocity is zero. The current dipole terms $m_{\BH} (\vec{v}_{\BH} \cdot \hat{k}) (\vec{r}_{\BH}\cdot  \hat{k})  D^{-2}$
are indeed present in the $g_{00}$, $g_{0i}$, and $g_{ii}$ components of the binary metric,
(see Eq.~(2.15) in \citealt{2000PhRvD..61l4013A}; and Eq.~(6.4) in \citealt{2009PhRvD..80l4039J}).
This leads to an orientation-averaged TOA correction
\begin{equation}\label{e:dtcd}
\delta t_{\rm cd} \sim \frac{m_{\BH} v r }{D^2} \frac{\sqrt{f T}}{2\pi f} =
30 \, m_3 D_{\pc}^{-2} f_8^{-5/6} T_{10}^{1/2}\,\ns\,.
\end{equation}
The next correction is the mass quadrupole,
\begin{equation}\label{e:dtmq}
\delta t_{\rm mq} \sim \frac{m_{\BH} r^2 }{D^3} \frac{\sqrt{f T}}{2\pi f} =
9 \, m_3 D_{\pc}^{-3} f_8^{-11/6} T_{10}^{1/2}\,\ns
\end{equation}
Now consider terms higher order in the masses. The leading order term here is the quadrupolar radiation term
$M_{\SMBH}m_{\BH}/(D r) \sim m_{\BH} v^2/D$. which leads to TOA residuals $\delta t_{\rm GW}$ of Eq.~(\ref{e:dt}).
All other terms in the metric are smaller by positive integer powers of the small parameters
in Eq.~(\ref{e:approx}).

In addition to the Einstein delay discussed above, the tidal effects of the binary
also induces epicyclic motion in the pulsar, which leads to variations in the path length to Earth.
This leads to a timing variation, $\delta t_{\rm R}$, analogous to the Roemer delay in pulsar binaries relative to the binary barycenter.
We estimate $\rp$, the
radius of the pulsar's epicyclic motion around the guiding center of its mean motion.
We assume that the pulsar exhibits oscillations on the forcing period, so that its angular velocity matches
the angular frequency of the forcing, $\omega$. The centripetal acceleration is then $\rp \omega^2$.
We consider the case where the forcing is due to a current dipole or a mass quadrupole,
$| F_{\rm cd,R}|/m_{\rm p} \sim m_{\BH} v r\sin \iota/D^3$ and
$|F_{\rm mq,R}|/m_{\rm p} \sim m_{\BH} r^2\sin \iota/D^4$.
The time lag is $\delta t_{\rm R} \sim 2 \rp\cos\mu$ in a half cycle, and
we assume the measurement improves with $(fT)^{1/2}$ for longer observations. Neglecting factors
of order unity, we get
\begin{align}\label{e:dtcdD}
\delta t_{\rm cd,R} &\sim \frac{m_{\BH} v r }{D^3} \frac{\sqrt{f T}}{(2\pi f)^2} =
5 \, m_3 D_{\pc}^{-3} f_8^{-11/6} T_{10}^{1/2}\,\ns\,,\\\label{e:dtmqD}
\delta t_{\rm mq,R} &\sim \frac{m_{\BH} r^2 }{D^4} \frac{\sqrt{f T}}{(2\pi f)^2} =
1 \, m_3 D_{\pc}^{-4} f_8^{-17/6} T_{10}^{1/2}\,\ns\,.
\end{align}
These estimates are consistent with that of \citet{2005ApJ...627L.125J}. They found
timing residuals corresponding to $\delta t_{\rm mq, R}$ can be 500\,ns
for a pulsar inside a globular cluster with $D_{\pc}=0.03$, $f_8=0.3$, $T_{10}=1$
for a $10\,\Msun$ BH orbiting around an $10^{3}\,\Msun$ IMBH.
The residuals are much larger in the near-field zone of the Galactic nucleus where the central object is a SMBH.

In conclusion, the Einstein delay terms $\delta t_{\rm cd}$ and $\delta t_{\rm mq}$ given by
Eqs.~(\ref{e:dtcd}--\ref{e:dtmq}) are much larger than the Roemer delay
$\delta t_{\rm cd,R}$ and $\delta t_{\rm mq,R}$ for $D\gtrsim\pc$ and $f>10\,\nHz$.
In particular, $\delta t_{\rm cd}$ is comparable to the standard radiative GW modulation,
$\delta t_{\rm GW}$ at $10\,\nHz$.
The gravitational near field timing residuals of individual sources
become much larger for pulsars at $D \sim 0.1\pc$,
at the level $\delta t_{\rm cd}\sim \delta t_{\rm mq,R}\sim 10\,m_3\,\mu$s at $f\sim10\,m_3\,\nHz$,
and even larger for smaller $f$. For stellar mass BHs, the individual contributions
of residuals are overlapping in Fourier space where near-field effects are most important
(see Eq.~(\ref{e:fsn})), and the net timing residual can be estimated by scaling
with $\sqrt{\D N/\D \ln f}$ using Eq.~(\ref{e:hccirc}).

\section{Results}
\begin{figure*}
\centering
\mbox{\includegraphics[width=8.7cm]{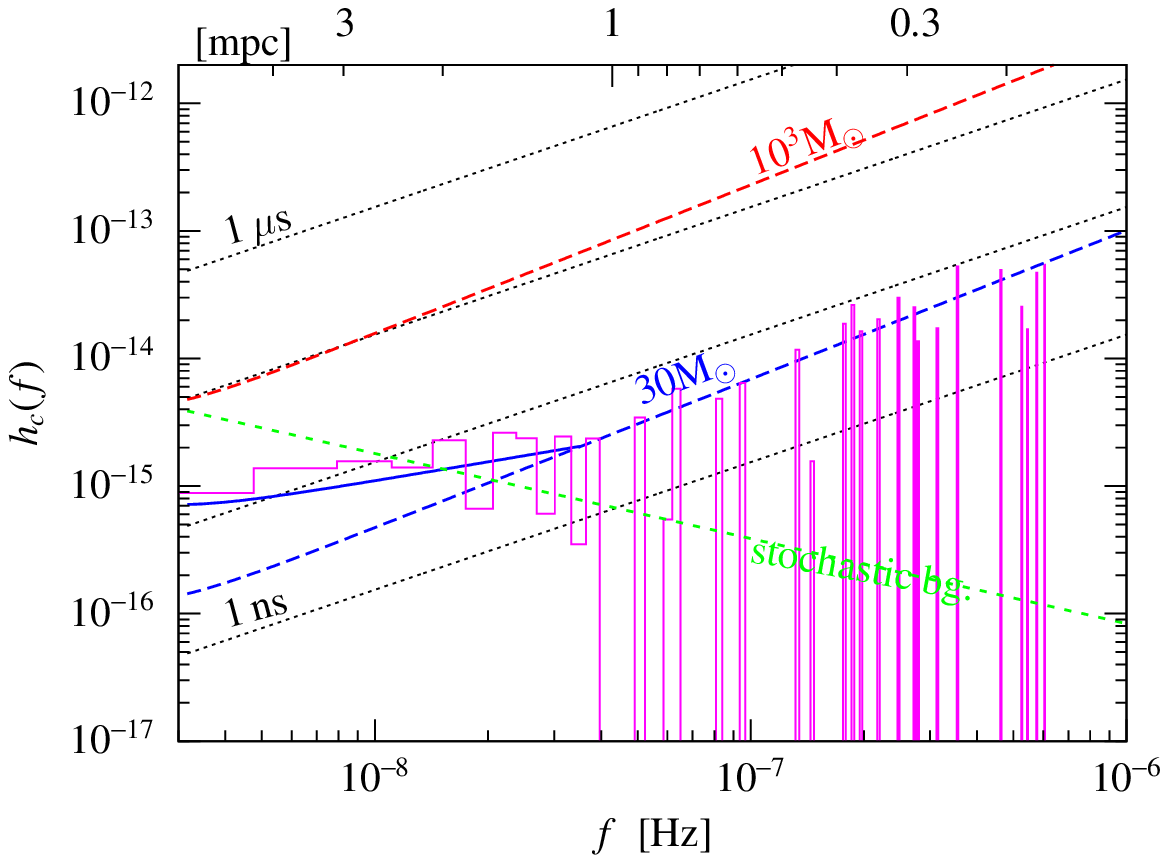}}
\mbox{\includegraphics[width=8.7cm]{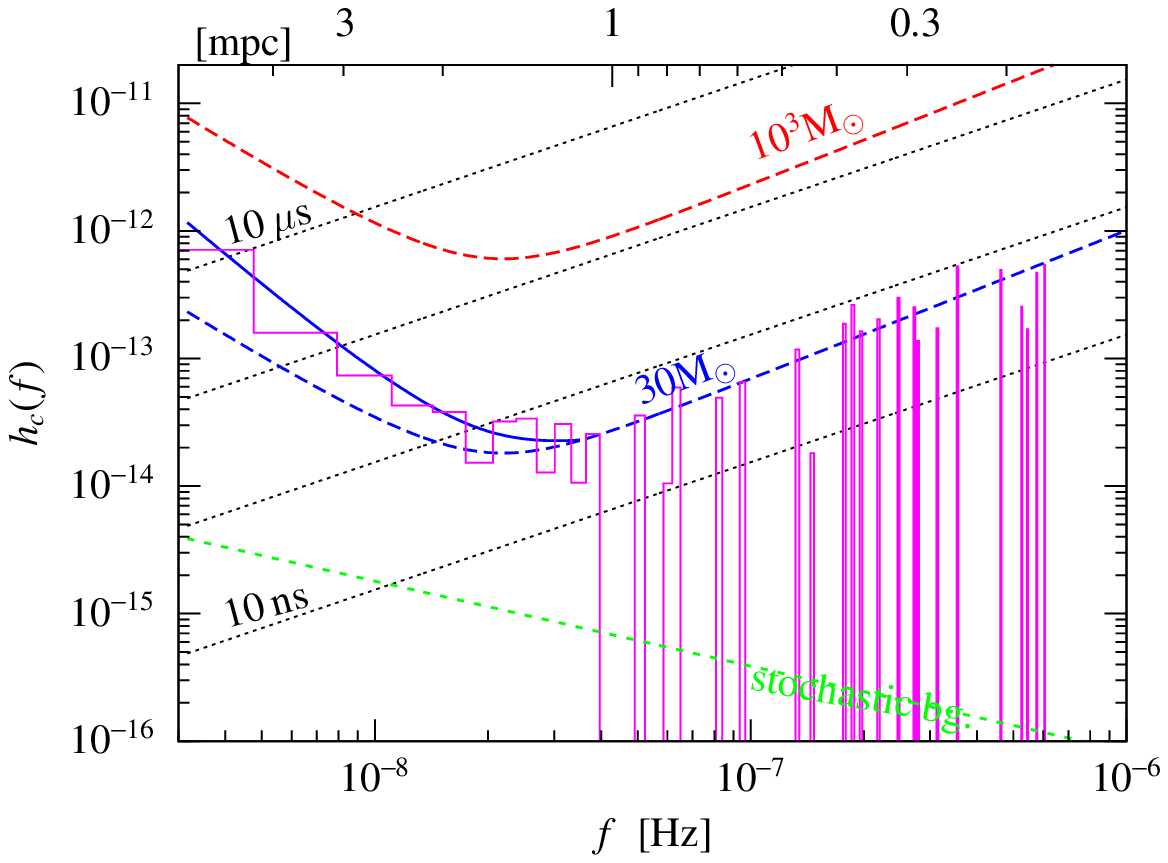}}
\caption{\label{f:hchar-alpha}
Same as Fig.~\ref{f:hchar} but for $1000$ BHs of mass $30\,\Msun$ within $1\,\pc$, with
a steeper density profile $r^{-2.5}$. There are more resolvable sources in this case.
}
\end{figure*}

We can now combine the results above to draw conclusions on the detectability of GWs from the
GC with pulsars in its neighborhood.
From Eq.~(\ref{e:dt}), the distance within which a PTA could measure the GWs of an individual source
with a fixed timing precision $\delta t=10 \, \delta t_{10}\, \ns$ is
\begin{equation}\label{e:Ddt}
D_{\delta t} = 14\, m_3 \delta t_{10}^{-1} T_{10}^{1/2} f_8^{1/6} \,\pc.
\end{equation}
Eqs.~(\ref{e:hci}) and (\ref{e:hcs}) show that
the GWs from an individual BH in the GCs rises above the stochastic GW background within a distance
\begin{equation}\label{e:Db}
D_{\rm bg} = 8.7\, m_3 T_{10}^{1/2} f_{8}^{11/6}  \,\pc\,.
\end{equation}
A pulsar within $D_{\delta t}$ and $D_{\rm bg}$ to the GC could be used to detect GWs from
individual objects in the GC. These estimates are conservative.
First, they assume an average orbital and pulsar orientation, $D_{\rm b}$ and $D_{\delta t}$ might be
2--3$\times$ larger for certain orientations. Second, Eq.~(\ref{e:Db})
assumes that small number statistics and gas effects are negligible in cosmological SMBH mergers,
and { might overestimate} the background \citep{2008MNRAS.390..192S,2011MNRAS.411.1467K}.\footnote{
{ Note that the signal from individual cosmological SMBHs can also be enhanced by a factor 2--3 for certain orientations,
but this effect is washed out for the unresolved stochastic background.}}

Figure~\ref{f:hchar} shows the orientation averaged characteristic GW amplitude for a 10 year
observation, incorporating the additional near-field effects using Eqs.~(\ref{e:hcPTA})
and (\ref{e:dtcd}--\ref{e:dtmqD}). The lower and top $x$-axis shows the GW frequency
and orbital radius for circular sources. The magenta curve displays the spectrum of timing residuals
for a Monte Carlo realization of a population of { 20,000 BHs within 1\,pc with number density $n(r)\propto r^{-2}$
and random orientation with mass  $10\Msun$ on circular orbits.}
The solid blue line shows the RMS
foreground of a cusp of stellar mass BH averaged over different realizations. The spectrum
separates into distinct spectral spikes at higher frequencies with RMS maxima shown by the dashed line.
 { Figure~\ref{f:hchar-alpha} shows the GW amplitude of a realization of
1000 BHs with mass $30\,\Msun$ in 1\,pc with a steeper number density profile 
$n(r)\propto r^{-2.5}$. Despite of the smaller overall number
of sources in the cluster, there are many more resolvable sources within 1 mpc in this case.}

The net background level in Figures~\ref{f:hchar} and \ref{f:hchar-alpha} is conservative,
as the background is sensitive to the RMS mass of objects in the cluster, which may
significantly exceed the mean.
Furthermore, individual sources may generate up to $\sim 3\times$ larger timing
residuals for certain orientations.
Timing pulsars at a distance $0.1$--$1\,\pc$ from the Galactic Center with $100\,\ns$--$10\,\mu$s timing precision
can be used to detect IMBHs of mass $10^3\,\Msun$, if they exist within $6\mpc$ of SgrA$^{*}$.
{ A population of stellar BHs within 2--5\,mpc generates timing variations greater than 100\,ns--10$\mu$s
for a pulsar within $0.1\pc$. However it will be much more difficult to
individually resolve $10\,\Msun$ stellar BHs within $1\mpc$ to SgrA$^*$, which
 requires an extreme 1--20\,ns timing precision.}

\begin{figure*}
\centering
\mbox{\includegraphics[width=8.7cm]{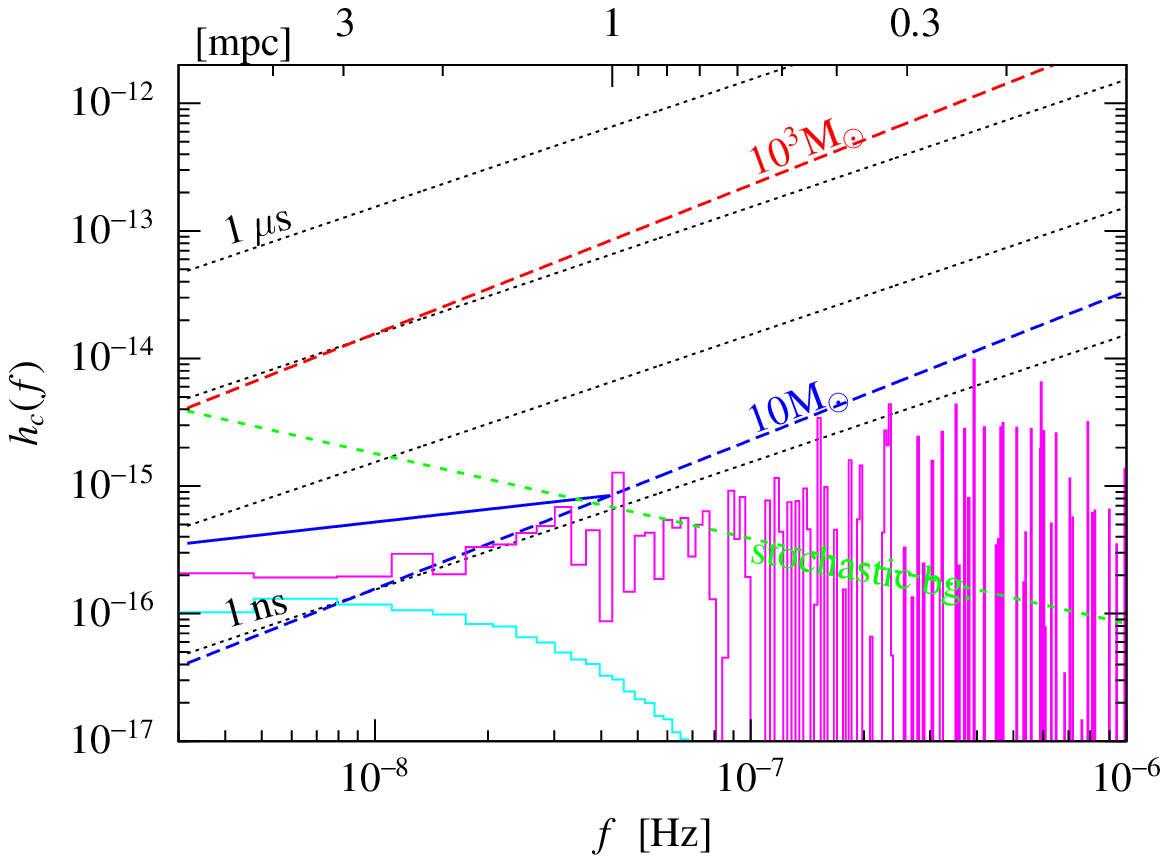}}
\mbox{\includegraphics[width=8.7cm]{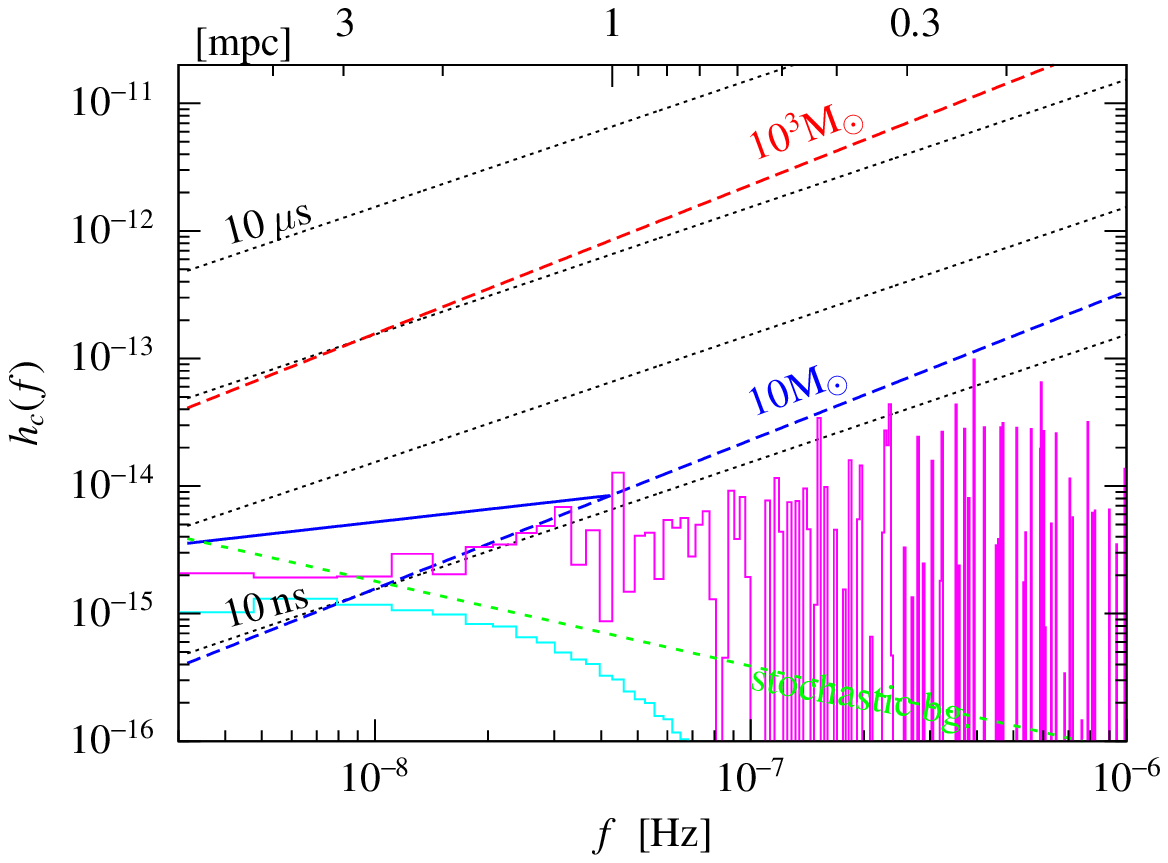}}
\caption{\label{f:hchar-ecc}
Characteristic spectral amplitude $h_c(f)$ for timing a
pulsar at 1\,pc (left) and 0.1\,pc (right) from the GC
for a population of $10\Msun$ BHs
with isotropic thermal eccentricity distribution. The contribution of periodic sources are shown with magenta,
burst sources with cyan. For comparison, the dashed blue and red lines show the RMS GW level
for individual circular sources, and solid blue lines show the RMS level from a population of circular
$10\,\Msun$ sources (see sections 2.1 and 2.2). { Note that} timing variations due to gravitational near field effects are not shown,
{ which dominate within $3\times 10^{-8}\,\Hz$ (see Figures~\ref{f:hchar} and \ref{f:hchar-alpha})}.
%\vspace{0.1cm}
}
\end{figure*}
Figure~\ref{f:hchar-ecc} shows the characteristic GW spectra for a Monte Carlo realization of an
eccentric population of $10\,\Msun$ BHs with an isotropic thermal eccentricity distribution.
Magenta and cyan lines show the contribution of periodic and burst sources, respectively. The dashed
lines show $h_c(f)$ for individual circular sources, and the solid blue lines represent the RMS GW level
for a population of circular sources for comparison.
The net GW spectrum of burst sources is typically less than the level of
periodic sources. The figure verifies the analytical calculations of Appendix~\ref{app:eccentric},
the continuous low-frequency spectrum of periodic sources is indeed comparable to the circular
level, modulo a weakly frequency dependent constant between $0.4\lesssim K\lesssim 0.8$.
Note that for clarity, we are not including the gravitational near field effects here, which
would dominate over the continuous low-frequency GW spectrum for
$D\sim 0.1\,\pc$ as shown in Figure~\ref{f:hchar}.

The transition to a spiky spectra happens at somewhat
larger frequencies for eccentric sources (c.f. Figs.~\ref{f:hchar}~and~\ref{f:hchar-ecc}).
Since individual sources generate GW spectra with many orbital upper harmonics,
the net high frequency spectrum is more complicated than in the circular case. The orientation averaged
spectral level for individual sources is typically less than $\delta t \lesssim 200\,\ns\, m_3 D_{\pc}^{-1}$
per frequency bin, however, the total root-sum-squared signal of all upper harmonics of individual sources is
much larger than the level of a single bin. In Appendix~\ref{app:eccentric} we show that the SNR
for detecting individual sources with a matched filter is comparable for eccentric and circular sources.
In this sense, the dashed lines in Fig.~\ref{f:hchar-ecc}, are representative of the total timing residual
of individual sources as a function of pericenter frequency for arbitrary eccentricity.
However, the full spectrum is rich in narrow features for eccentric sources and
pericenter precession slowly modulates the amplitude
of the timing residuals in individual pulsars. Both of these features could help to separate
individual eccentric sources from the signal of other sources in GC.

\section{Discussion}\label{s:discussion}
We have shown that pulsars within a few pc distance from the GC offer a unique
probe to identify IMBHs and stellar BHs orbiting around SgrA$^*$.
An IMBH, if present in the GC, sinks to orbital radii corresponding to the pulsar
timing frequency bands.
Depending on the binary orientation, orbital frequency, and pulsar distance,
the GWs and gravitational near-field effects modulate the TOAs by a few ns to
$100\,\mu$s
for these sources with masses between $10\,\Msun$  and $10^4\,\Msun$. Based on the GW spectral features, the signals of more than
one IMBH (up to tens, if present) could be individually resolved and isolated from the fainter
signal of stars and stellar mass objects and the cosmological GW background.

This observational probe is complimentary to EM measurements with different systematics.
GWs are generated by all gravitating objects, including those that are black and undetectable
in EM bands. GWs escape the galactic nucleus without any dissipation or dispersion.
Closer to SgrA$^*$, sources generate
stronger, higher frequency GWs, the number of observable cycles increases,
and the number of objects per frequency bin decreases. Thus,
unlike EM imaging techniques, the prospects for detecting and resolving individual
objects through GW measurements improve closer in towards SgrA$^*$,
even if the number density of objects increases inwards steeply.
Furthermore, the gravitational effects are proportional to the RMS mass of objects,
making the measurement more sensitive to individual higher mass objects in the distribution, even
if the total mass of lighter objects is somewhat larger on comparable radius orbits.

Repeated pulsar timing observations over a few year baseline could
reveal a detailed census of BHs in the inner mpc of the GC.
We have shown that the
GW foreground of stellar BHs rises above the cosmological background and separates into distinct
peaks above $40\,$nHz, corresponding to a GW period of less than 1\,yr or orbital separations
less than 1\,mpc. Based on a simple estimate using circular orbits, we found that the
total number of individually resolvable BHs is between 7--70, depending on the
radial number density distribution exponent $r^{-\alpha}$ between $r^{-1.75}$ and $r^{-2.25}$.
These observations are therefore exponentially sensitive to $\alpha$, capable of testing the
theory of strong mass segregation \citep{2009ApJ...698L..64K}.

Eccentricity complicates the spectral shape of resolvable
sources by adding upper frequency harmonics. Although the timing residuals in individual
frequency bins is suppressed by this effect relative to circular orbits, the total timing residual SNR with
a matched filter is comparable for eccentric sources.  At lower frequencies,
a population of objects on eccentric orbits generates a stochastic GW foreground
with a similar spectral shape and a comparable amplitude,
as a circular population.
Objects on larger-semimajor axis, eccentric orbits generate GW bursts during pericenter passage near SgrA$^{*}$.
The stochastic GW burst signal is much less than the level of periodic sources.

This analysis hinges on the assumption that future surveys will discover pulsars near the GC that can be timed to the sufficient accuracy.
Most of the observed S-stars within a few mpc and the young O/B stars in the GC
will eventually turn into pulsars in a supernova explosion \citep{2004ApJ...615..253P}.
Based on the age and number of these stars, and assuming that we don't live in a special time,
one might expect more than $10^4$ pulsars in the GC.
Some of these may become MSPs, and may be beamed towards us to be
detectable with future SKA-type instruments \citep{2004NewAR..48.1413C}.
They might be expected to segregate to the outskirts of the GC on a Gyr timescale
as heavier objects sink inwards \citep{2002ApJ...571..320C}.
{ Recently, \citet{2012ApJ...747....1L} examined the expected timing accuracy of pulsars in the GC
accounting for radiometer noise, pulse phase jitter, and the interstellar scintillation of the ISM.
They found that the $1\,\hr$ timing accuracy of SKA is expected to be between $10$--$100\,\mu{\rm s}$
for regular pulsars. Our results indicate that the necessary accuracy to detect timing variations associated to individual
$10\,\Msun$ BHs within 1\,mpc requires much higher timing accuracy, which might be prohibitively difficult
even with MSPs with a factor of 100--1000 better timing accuracy. However, the net variations caused by a population of these objects is
detectable between 2--5 mpc at these accuracy levels. Remarkably,  a $10$--$100\,\mu{\rm s}$ timing accuracy is
sufficient to individually resolve or rule out the existence of $10^3\,\Msun$ IMBHs within 5 mpc from SgrA*.}

As the GW spectrum is rich in strong spiky features at high frequencies, it
may be possible to isolate GW induced timing residuals from other systematic effects.
One such effect is if the pulsar itself is a part of a binary system.
Fortunately, however, binaries with orbital periods of years, matching the GW foreground of the GC,
are very soft, and are not long lived near the GC, they are easily disrupted by three-body encounters.
Indeed, due to the high velocity dispersion in the GC, $\sigma\sim 200\km/\sec$ at $D\sim\pc$, stellar mass
binaries are soft for orbital frequencies $f\lesssim (2\pi)^{-1} \sigma^3/(\G \Msun) \sim 9,600\,\nHz$,
or orbital period $f^{-1}\gtrsim 1.2\,$day.
The evaporation timescale on which a series of more distant encounters gradually increases the
binary separation, is $t_e\sim 0.06\,\sigma/(\G\rho a \ln \Lambda)\sim 40 f_8^{2/3}\,$Myr, where $\Lambda$ is the Coulomb logarithm, and
the ionization timescale to disrupt the binary by a close three body encounter is
$t_{i}\sim 0.04\, \sigma/(\G\rho a)\sim 70 f_8^{2/3}\,$Myr (see \S~7.5.7. in \citealt{2008gady.book.....B}).
Here we have expressed the binary semimajor axis $a$ with the orbital frequency, which is of order
$0.1\lesssim f_8\lesssim 10$ to match the GW signal. Thus,
MSPs which form in short period binaries, like in typical LMXBs, and become wide,
may be expected to typically become single in the GC. However, a few pulsar-BH binaries may form through
three body exchange interactions and may be longer lived in the GC \citep{2011MNRAS.tmp..911F}.
Ultimately, multiple pulsars would be necessary to rule out systematic effects.

GW observations with pulsar timing in the GC  could be combined with other observable channels to
map the GC. An IMBH orbiting around the SMBH would be observable with future millimeter
VLBI imaging with EHT \citep{2011ApJ...735...57B}. Conversely, if those observations reveal an IMBH,
the inferred orbital parameters could help in identifying the GW counterpart with pulsar timing.

\acknowledgments
BK acknowledges support from NASA through Einstein Postdoctoral
Fellowship Award Number PF9-00063 issued by the Chandra
X-ray Observatory Center, which is operated by the Smithsonian
Astrophysical Observatory for and on behalf of the National
Aeronautics Space Administration under contract NAS8-03060.
AR thanks the Director and members of the Institute for Theory and Computation (ITC),
Harvard University for hospitality. He also thanks Woldek Kluzniak, Duncan Lorimer
and Maura McLaughlin for discussions.
This work was supported by the Hungarian Research Fund  OTKA (grant 68228),
Netherlands Research Council NWO
(grants \#639.073.803), the Netherlands Research School for
Astronomy (NOVA), the Eleventh Five Year Plan Project No. 11P-409 at
Tata Institute of Fundamental Research (TIFR) and by a Visiting Scholarship at
the ITC.

\appendix
\section{Eccentric sources}\label{app:eccentric}
Here we present the mathematical derivation of the GW foreground of eccentric sources in the GC.
We start by reviewing the eccentric waveforms and GW spectra, then calculate the GW background
of an eccentric population of periodic sources and burst sources, respectively,
and finally discuss simple estimates of the signal-to-noise ratio of timing residuals
for individual eccentric sources.

\subsection{Waveform}

An individual eccentric source with semimajor axis $a$ and eccentricity $e$,
generates a GW strain composed of discrete upper harmonics with frequency $f_n=n\forb$,
\begin{align}
h(a,e,t) = \sum_{n=1}^{\infty} h_n(a,e;f_n) e^{2\pi i f_n t}
\end{align}
where
\begin{equation}\label{e:hn}
h_n(a,e;f_n) = \frac{2}{n}\sqrt{g(n,e)}\, h_0(a),
\end{equation}
where $h_0(a) = \sqrt{32/5}\, M_{\SMBH}m_{\BH}/(D a)$ is the GW strain amplitude for circular orbits, and
\begin{equation}\label{e:g}
 g(n,e)=\frac{n^4}{32}\left[\left(J_{n-2}- 2e J_{n-1}
  + \frac{2}{n}J_n + 2 e J_{n+1} - J_{n+1}\right)^2
  +(1-e^2) (J_{n-2} - 2 J_n + J_{n+2} )^2 + \frac{4}{3n^2}
  J_n^2\right]\,.
\end{equation}
Here $J_i\equiv J_i( x)$ is the $i$th Bessel function evaluated at $x=ne$ \citep{1963PhRv..131..435P},
and we have RMS averaged over the binary inclination.
The Fourier transform of $h(a,e,t)$ measured for some time $T$ is then
\begin{align}\label{e:tildeh}
\tilde{h}(a,e,f) = \sum_{n=1}^{\infty} h_n(a,e;f_n)\, T\, w(f,f_n,T)\,,{\rm~where~~} w(f,f_n,T)=\frac{\sin[2\pi (f- f_n)T]}{2\pi (f-f_n)T}\,.
\end{align}
Here $w(f,f_n,T)$ is the Fourier transform of a window function of width
$T$ and a unit integral.
It approaches unity at $f=f_n$ and cuts off for $|f-f_n| \gtrsim 1/2T$. The spectral width
of each GW harmonic is $\Delta f \sim 1/T$.

For circular sources, $g(n,0)=1$ for $n=2$ and 0 otherwise. For eccentric sources,
the dominant frequency harmonic is $\fp = \np \forb$, the inverse pericenter passage timescale,
\begin{equation}\label{e:fp}
\np(e)
= {\rm ceil}\left[1.15 \frac{(1+e)^{1/2}}{(1-e)^{3/2}} \right]\,,{\rm ~~so~that~~~}
\fp(a,e)\sim \frac{(1+e)^{1/2}}{(1-e)^{3/2}} \forb(a)\,,
\end{equation}
where $\rm ceil(x)$ is the nearest integer larger than $x$ \citep{2009MNRAS.395.2127O}.
The emitted GW spectrum is broadband with a maximum near $\fp$, where $90\%$ and $99\%$ of the power is
between $0.2\, \fp < f < 3\, \fp$ and between $0.1\,\fp < f < 5\, \fp$, respectively.
The $g(n,e)$ harmonic weights in Eq.~(\ref{e:g}) have a maximum near $n_{p}(e)$.

The definition of the inclination-averaged GW strain amplitude (\ref{e:hn}) can be made more lucid
by recalling the definition of the GW flux, $S=\dot h(t)^2/16\pi$, and verify if the total power output is consistent with
Eq.~(16) of \citet{1963PhRv..131..435P}. Indeed,
\begin{equation}\label{e:P}
P = 4\pi D^2 S = \frac{1}{4}D^2 \dot h^2 = \pi^2 D^2 \sum_{n=1}^{\infty} n^2 \forb^2 h_n^2
=4\pi^2 D^2 \forb^2 h_0^2 F(e) = \frac{32}{5} \frac{M_{\SMBH}^3 m_{\BH}^2}{ a^5} F(e)\,
\end{equation}
where we have used Kepler's law $\omega_{\rm orb}^2 = 4\pi^2 \forb^2 = M_{\SMBH}/a^3$, the definition of $h_0$,
\begin{equation}
F(e)\equiv \sum_{n=1}^{\infty} g(n,e) = \frac{F_1(e)}{ (1- e^2)^{7/2} }\,,{\rm~and~~} F_1(e)=1 + (73/24) e^2 + (37/39) e^4\,.
\end{equation}

\subsection{GW background of periodic sources}
Let us estimate the net contribution of many sources to the GW background if observed for time $T$.
For a source with semimajor axis $a$ and eccentricity $e$, the counts in each frequency bin $f$ are given by
\begin{equation}\label{e:hc1}
 h_{c,1}^2(a,e;f) =
 \tilde{h}^2(a,e;f)\, f\, \Delta f =
\sum_{n=1}^{\infty} f T\,h_0^2(a)\, \frac{4}{n^2} g(n,e)
\times\left\{
\begin{array}{ll}
1 &~~{\rm if}~ | f - f_n(a) | \leq \Delta f/2~~{\rm and~} f T>1\\
0 &~{\rm otherwise}
\end{array}
\right.\,.
\end{equation}
where $f_n(a) = n\forb(a) = n(2\pi M_{\SMBH})^{-1}\,  (a/M_{\SMBH})^{-3/2}$ is the GW frequency of the $n^{\rm th}$
upper harmonic for a fixed semimajor axis.
Conversely, the range of semimajor axis, $a_n\pm 0.5\Delta a_n$,
for which the $n^{\rm th}$ harmonic contributes to the frequency bin between $f\pm 0.5\,\Delta f$, is
\begin{equation}\label{e:an}
 a_n = M_{\SMBH} (2 \pi M_{\SMBH} f/n)^{-2/3} = \left(\frac{n}{2}\right)^{2/3}a_2\,.
\end{equation}
where $\Delta a_n = |\D a_n /\D f|\, \Delta f= \frac{2}{3} a_{n}/(f T)$.

Now let us assume a phase space distribution, in which the number of objects in the neighborhood of $(a,e)$ is
 $\D^2 N = (\partial^2 N/ \partial a\partial e)\, \D a \D e$. The number of sources that contribute to the $n^{\rm th}$ harmonic is
$\Delta N_n = \int \D e (\partial^2 N/ \partial a_n\partial e)\, \Delta a_n$. The GW signal of all sources is then
\begin{equation}\label{e:hc3}
 h_{c, {\rm per}}^2(f) =
\int_0^1 \D e\;\sum_{n=1}^{n_{\max}}  \left[\left(\frac{\partial^2 N}{\partial a\partial e} \left|\frac{\D a}{\D f}\right|\Delta f\right)
 h_{c,1}^2(a,e;f)\right]_{a=a_n}\,.
\end{equation}
For periodic sources, $n_{\max}$ is set by the condition that objects are observed for at least one orbit,
$1/f_{\rm orb}(a_n)=n/f\lesssim 1/T$, implying that $n_{\max}=f T$.
We shall consider the contribution of burst sources, on larger radius orbits observed for only a fraction of the orbit, separately below.
Rearranging and using Eq.~(\ref{e:hc1}),
$(\partial^2 N/ \partial a\partial e) = 4\pi a^2 n_{\BH}(a) \varphi(e)$, and $\D a/ \D f = \frac{2}{3}a/f$,
gives
\begin{equation}
 h_{c, {\rm per}}^2(f) =
%\int_0^1 \D e\;\sum_{n=1}^{\infty}  4\pi  a_n^2\,  n_{\BH}(a_n) \,\varphi(e)\,\left(\frac{2}{3}\frac{a_n}{f} \right)\,f\,  h_0^2(a_n) \,\frac{4}{n} g(n,e)=
 \sum_{n=1}^{fT}\frac{8\pi}{3} a_n^3\,  n_{\BH}(a_n)\,h_0^2(a_n) \int_0^1 \D e\; \,\varphi(e)\, \frac{4}{n^2} g(n,e)\,.
\end{equation}
Let us assume that $n_{\BH}(a) \propto a^{-\alpha}$. Then using
$h_0(a)\propto a^{-1}$ and Eq.~(\ref{e:an}),
we find that the RHS is proportional to $a_n^{1-\alpha}$. Now let us express $a_n$ with $a_2$ using Eq.~(\ref{e:an}),
\begin{equation}\label{e:hc2}
 h_{c, {\rm per}}^2(f) = \frac{8\pi}{3}  a_2^3\,  n_{\BH}(a_2)\,h_0^2(a_2)\, K_{\rm per}(f)
\end{equation}
where
\begin{equation}\label{e:Kper}
 K_{\rm per}(f)= \int_0^1 \D e\; \sum_{n=1}^{fT} \left(\frac{2}{n}\right)^{(2+\alpha)2/3} \,\varphi(e)\,  g(n,e)\,.
\end{equation}
Thus $K_{\rm per}(f)=0$ at $f\leq 1/T$, and increases monotonically, and asymptotes a constant for  $n_{\max}=fT \rightarrow \infty$. One
can show\footnote{To see this, consider an approximate sharply peaked signal around $\fp$, for which
$g(n,e)\sim \delta_{n,n_p(e)} F(e)$, where $\delta_{ij}=1$ if $i=j$ and 0 otherwise.}
that this constant is insensitive to the highest eccentricity sources if $\alpha> 1/2$, which is expected to be satisfied.
For $\alpha=2$ and a thermal distribution of eccentricities $\varphi(e)=2e$, we get $K_{\rm per}\sim 0.8$ for $fT\gg 1$. This together
with Eq.~(\ref{e:hc2}) shows that the net GW spectrum of a continuous population of eccentric sources is very similar to
that of circular sources. However, the signal is much different for individually resolvable sources,
as they are comprised of many upper harmonics

\subsection{GW background of burst sources}
GW bursts are generated by objects which make only one close approach near the SgrA* during the observation. These sources
are on eccentric orbits with orbital time $\forb^{-1}$ exceeding $T$, but for which the pericenter timescale $\fp^{-1}$ is less than $T$
and the orbital phase is such that pericenter passage occurs within the observation. The later condition means that only a
$1/(\fp T)$ fraction of all such sources will contribute in the observation time. For a fixed measurement frequency, $f$, therefore
$\forb=f/n<1/T$, $f \np/n > 1/T$. Therefore $f T < n < \np f T$ and the fraction among these sources
that contribute is $n/(\np f T)$.

Repeating the derivation for the net GW burst background over time $T$, Eq.~(\ref{e:hc1}--\ref{e:hc3}), we get
\begin{equation}\label{e:hcburst}
 h_{c, {\rm b}}^2(f) = \frac{8\pi}{3}  a_2^3\,  n_{\BH}(a_2)\,h_0^2(a_2)\, K_{\rm b}(f)
\end{equation}
where
\begin{equation}\label{e:Kb}
 K_{\rm b}(f)= \int_{e_{\min}}^1 \D e\; \sum_{n=fT}^{n_p f T} \frac{n}{n_p f T} \left(\frac{2}{n}\right)^{(2+\alpha)2/3} \,\varphi(e)\,  g(n,e)\,.
\end{equation}
Note that resolving the burst source also requires that the sampling frequency $f_{\max}=2/\Delta t$ and timescale between
observations $\Delta t$ to satisfy $\fp\lesssim f_{\max}$ and $\Delta t/2 \lesssim \fp^{-1}$.
This implies that $n_{\min}\gtrsim \frac{1}{2} n_p f \Delta t$ has to hold.
This requirement, however, is typically already satisfied for most sources if $\Delta t\gtrsim 1\,$week, the number of sources on so
eccentric orbits with such small $\rp$, is close to zero. Comparing $K_{\rm per}(f)$ and $K_{\rm b}(f)$, Eqs.~(\ref{e:Kper}) and (\ref{e:Kb}),
shows that the net GW signal of periodic sources exceeds the contribution of GW burst sources.

\subsection{Signal to noise ratio}
Here we present simple estimates on the scaling of the GW signal and timing SNR for eccentric sources
using elementary functions. This is useful not only because the exact signal presented above is algebraically complicated,
but also because the total SNR for individual sources is not represented well by $h_c(f)$ in a single frequency bin,
but it is sensitive to the coherent sum over many orbital harmonics.

Let us define an effective GW strain amplitude using that the GW power at pericenter passage, $P_{p}$, is the average power
times the fraction of time the source spends near pericenter passage, i.e. $n_p$, given by Eq.~(\ref{e:fp}).
Thus,
\begin{equation}
 P_{p} = \frac{1}{4} D^2 \omega_p^2 h_p^2 = n_p P =  \frac{32}{5} \frac{M_{\SMBH}^3 m_{\BH}^2}{ \rp^5} (1+e)^{1/2}
F_1(e) \,.
\end{equation}
This amounts to the root-sum-square of individual frequency harmonics in the spectrum for single sources, without
averaging over the full orbit (c.f. Eq.~(\ref{e:P})).
From this we get, that the effective strain at close passage is
\begin{equation}
 h_p = 2\frac{(n_pP)^{1/2}}{D\, \omega_p}  =   \sqrt{\frac{32}{5}} 2\frac{M_{\SMBH} m_{\BH}}{D\, \rp}
F_1^{1/2}(e)\,
\end{equation}
where $\omega_p = 2\pi f_p$ and we have used $\omega_p = M^{1/2} (1+e)^{1/2} \rp^{-3/2}$.
The pulsars are sensitive to the time integral of the strain (see \S~\ref{s:ptas}),
\begin{equation}\label{e:Hc}
{\cal H}_c = \frac{h_p}{\omega_p} (f_{\orb} T)^{1/2}
= \sqrt{\frac{32}{5}} \frac{1}{\pi} \frac{M_{\SMBH}^{3/4} m_{\BH}}{D a^{1/4} } T^{1/2} \frac{(1-e)^{1/2}}{(1+e)^{1/2}}
F_1^{1/2}(e)\,.
\end{equation}
The SNR is then $S/N = {\cal H}_c/\delta t_0$ where $\delta t_0$ is the timing noise over time $\fp^{-1}$.
Note that the RMS of the eccentricity dependent terms is 0.438 in Eq.~(\ref{e:Hc}),
if the eccentricity is drawn from a thermal distribution $\varphi(e) = 2\,e$ between $0\leq e \leq 1$  \citep{2008gady.book.....B}.
The SNR of the timing residual is not very sensitive to the semimajor axis, or the maximum eccentricity in the cluster.
For fixed $a$, eccentric sources contribute to the net timing residuals at a similar level as circular sources.

Eq.~(\ref{e:Hc}) is only applicable for periodic sources, i.e. if $f_{\rm orb} T \geq 1$.
For a single GW burst source  with $1/f_{\rm p}<T<1/f_{\rm orb}$,
\begin{equation}
{\cal H}_{c,1\rm b} = \frac{h_p}{\omega_p}
=  \frac{1}{2\pi}\sqrt{\frac{32}{5}} \frac{M_{\SMBH}^{1/2} m_{\BH}}{D }  \frac{\rp^{1/2}}{(1+e)^{1/2}}
F_1^{1/2}(e)\,,
\end{equation}
provided that the orbital phase coincides with pericenter passage during the observation.
This shows that for burst sources with fixed pericenter distance,
the total timing residual is weakly dependent on eccentricity,
for fixed pericenter distance. The number of sources is constant as a function of $\rp$ for fixed $e$, but the
fraction of sources that are near pericenter passage at any given instant  decreases quickly for smaller $\rp$ proportional to
$1/n_p$. Therefore the net contribution of burst sources scales as
\begin{equation}
{\cal H}_{c,\rm b} =\Delta N {\cal H}_{c,1\rm b} =  {\cal H}_{c,1\rm b} (\omega_p T )^{1/2}  n_{rp}\Delta \rp = \frac{h_p}{\omega_p} (\omega_p T )^{1/2} n_{rp} \Delta \rp
=  \frac{1}{\pi}\sqrt{\frac{32}{5}} \frac{M_{\SMBH}^{3/4} m_{\BH}}{D \rp^{1/4}}  \frac{T^{1/2} n_{rp} \Delta \rp}{(1+e)^{1/2}}
F_1^{1/2}(e)
\end{equation}
The spectral density within a frequency bin follows from $\Delta r_{p} = (2/3) \omega_p^{-5/3}\Delta \omega_p$
implying that ${\cal H}_{c,u} \propto \rp^{-1/4}\Delta \rp \propto \omega_p^{-3/2} \Delta \omega_p$. The average
timing spectral density of burst sources decreases quickly toward higher frequencies.

\bibliographystyle{apj}
\bibliography{apj-jour,p}

\end{document}